# SLM-based Active Focal-Plane Coronagraphy: Status and future on-sky prospects

Jonas G. Kühn*[a], Laurent Jolissaint[b], Audrey Bouxin[b], and Polychronis Patapis[c]

[a]Space Sciences Institute, University of Bern, Gesellschaftsstrasse 6, 3012 Bern, SWITZERLAND;
[b]University of Applied Sciences HEIG-VD, Route de Cheseaux 1, 1401 Yverdon-les-Bains, SWITZERLAND;
[c]Institute for Particle Physics and Astrophysics, ETH Zurich, 8093 Zurich, SWITZERLAND

## ABSTRACT

We recently started to investigate how liquid-crystal on silicon (LCOS) spatial light modulator (SLM) would perform as programmable focal-plane phase mask (FPM) coronagraphs. Such "adaptive coronagraphs" could potentially help adapt to observing conditions, but also tackle specific science cases (e.g. binary stars). Active FPMs may play a role in the context of segmented telescope pupils, or to implement synchronous coherent differential imaging (CDI). We present a status update on this work, notably early broadband contrast performance results using our new Swiss Wideband Active Testbed for High-contrast imaging (SWATCHi) facility. Finally, we unveil the upcoming near-infrared PLACID instrument, the Programmable Liquid-crystal Adaptive Coronagraphic Imager for the 4-m DAG observatory in Turkey, with a first light planned for the end of the year 2022.

## 1. INTRODUCTION

Novel principles of "active coronagraphy" have recently started to gain momentum in the exoplanets high-contrast imaging instrumentation community, notably because most upcoming large telescope apertures – whether ground-based or space observatories – will rely on segmented primary mirrors, whose merit figures may detrimentally evolve over the lifetime of the facilities. Additional potential benefits of such "active" or "adaptive" coronagraph schemes include the ability to (i) improve observational efficiency by optimizing the coronagraphic phase or amplitude pattern in function of environmental conditions (seeing, wind speed etc.), (ii) to null resolved giant stars or multiple star systems,[1] (iii) to ease the execution of built-in self-calibration procedures (e.g. for non-common path errors wavefront sensing), and (iv) to implement promising coherent-differential imaging (CDI) detection methods.[3] Among the proposed active coronagraphy approaches is to use liquid-crystal on-silicon spatial-light modulators (LCOS-SLM) display panels, acting as pixelated programmable focal-plane phase mask (FPM) coronagraphs.[1-3] Indeed, the micron-scale of the those SLM panels, combined with the typically large CMOS chip size (> 1 Mpx) makes this technology attractive for focal-plane applications, where one needs to sample the point-spread-function (PSF) with Nyquist resolution or better. Further, those panels can run at video-rate or better (up to 700 Hz), and are available off-the-shelves with full-wave phase stroke up to the telecom band (astronomical H-band). We however note that pupil-plane applications of this technology have also recently gathered interest, particularly to be able to use the high pixel counts and refresh rate to be able to simulate atmospheric turbulences in the laboratory.[4] On the downside, the LCOS-SLM approach suffers from non-negligible flaws in the coronagraphy context, particularly the limited throughput (strictly < 50pc) due the requirement of linearly-polarized light for phase-only modulation, as well as from the intrinsically scalar nature of the SLM phase modulation, as opposed to e.g. vectorial phase shift,[5,6] potentially limiting performance in broadband light. Nevertheless, in practice, it has to be noticed that most coronagraphs currently equipping ground-based high-contrast instruments are usually limited in contrast performance to a common "bottleneck", dominated by residual adaptive-optics (AO) wavefront errors, non-common path aberrations (NCPAs), and non-optimal telescope pupil shape (central obscuration, support spiders).

*jonas.kuehn@space.unibe.ch

Until recently, we only demonstrated FPM coronagraphs performances of commercially-available LCOS SLMs in monochromatic light conditions, in the visible ($\lambda$ = 633 nm), with typical achievable coronagraphic null depths in the order of 1.5·10$^{-2}$.[3] However, since 2018, efforts were initiated to build an all-reflective broadband SLM testbed, called SWATCHi (Swiss Wideband Active Testbed for Coronagraphic High-contrast imaging). The current version of the SWATCHi testbed is presented in Section §2, along with early insights on a few contrast measurement results in broadband visible light conditions.

Up until now, the advent of a SLM-based high-contrast imaging facility on a large ground-based telescope was admittedly in the domain of science-fiction, but this is about to change. Indeed, Section §3 introduces the PLACID (Programmable Liquid-crystal Active Coronagraphic Imaging for the DAG telescope) instrument concept, that is foreseen to come online by the end of 2022 on the upcoming Turkish National Observatory DAG 4-m telescope.[7,8] PLACID will consist in a fore-optics intermerdiate-stage high-contrast instrument, located in-between the TROIA extreme AO based on a pyramid wavefront sensor (WFS),[9] and the DIRAC near-infrared (NIR) focal-plane array (FPA) developed by the Australian AAO consortium, based around a HAWAII-2RG. PLACID will operate in the astronomical H-band (1.6 μm) as contractual baseline, but will equip a custom NIR "digital coronagraph" SLM actually in principle capable to operate up to the astronomical Ks-band (2.15 μm). The fact that the DAG telescope will field a high-contrast instrument such as PLACID, is foreseen to provide the nascent Turkish astronomical community with an easy-to-use versatile exAO exoplanet imaging capability, with large future potential for international collaborations ("on-sky laboratory" for prototyping of new coronagraphy concepts) and for testing of novel technological developments (polarization-insensitive SLMs, time-domain imaging etc.).

## 2. EARLY LAB CONTRAST RESULTS IN BROADBAND VISIBLE LIGHT

### 2.1 The SWATCHi SLM testbed

SWATCHi stands for the Swiss Wideband Active Testbed for Coronagraphic High-contrast imaging, which was initially build at ETH Zurich, Switzerland, before being recently relocated at the Space Sciences Institute of the University of Bern, Switzerland. The SWATCHi testbed (Fig.1) is an all-reflective (protected silver coated mirrors) high-contrast imaging testbed designed to test the coronagraphic contrast performance of various LCOS SLM panels that are operating as reflective FPMs (i.e. placed to operate in reflection in an intermediate focal-plane), under broadband light conditions. Although the SWATCHi bench is only operating in the visible light regime at the moment, due to the limited NIR sensitivity of its CCD camera (pco.pixelfly), it is otherwise essentially "NIR ready". Indeed, the setup operates with a versatile super-continuum light source (NKT Photonics SuperK Compact, 450-2400 nm spectral range) coupled to an intermediate stage fielding a bandpass filter wheel, with the outgoing light being again fiber-fed up to the SWATCHi entrance focal-plane (see Fig.1). An intermediate entrance pupil-plane (Fig.1) enables to insert various pupil masks (1-mm thick molybden laser cut) to simulate various telescope apertures at range of focal-ratios (at F/30 or slower on the SLM). As shown on Figure 1, the beam f-ratio is then slowed down by a factor 1.5x before reaching the coronagraphic reflective focal-plane, where the LCOS SLM panel is inserted (with a wire grid polarizer placed in front, in double-path configuration to mitigate internal reflection effects). The geometric beam configuration in this plane plays a crucial role for the phase modulation: depending of the SLM pixel pitch (usually < 10 μm), the f-ratio has to be adjusted to ensure Nyquist or better spatial sampling of the PSF (as a general rule a sampling of 10 pixels per $\lambda$/D is recommended to be able implement complex FPM patterns), and an off-axis reflection angle < 5 deg is advised to minimize inter-pixel crosstalk. An additional Lyot pupil-plane is then located downstream, to be able to insert Lyot masks (same manufacturing technique), before reaching the scientific focal-plane. Additionally, the last fold mirror is installed on a magnetic plate, enabling to mechanically switch to a pupil-imaging lens configuration, always useful for alignment and diagnostics (see Fig.1).

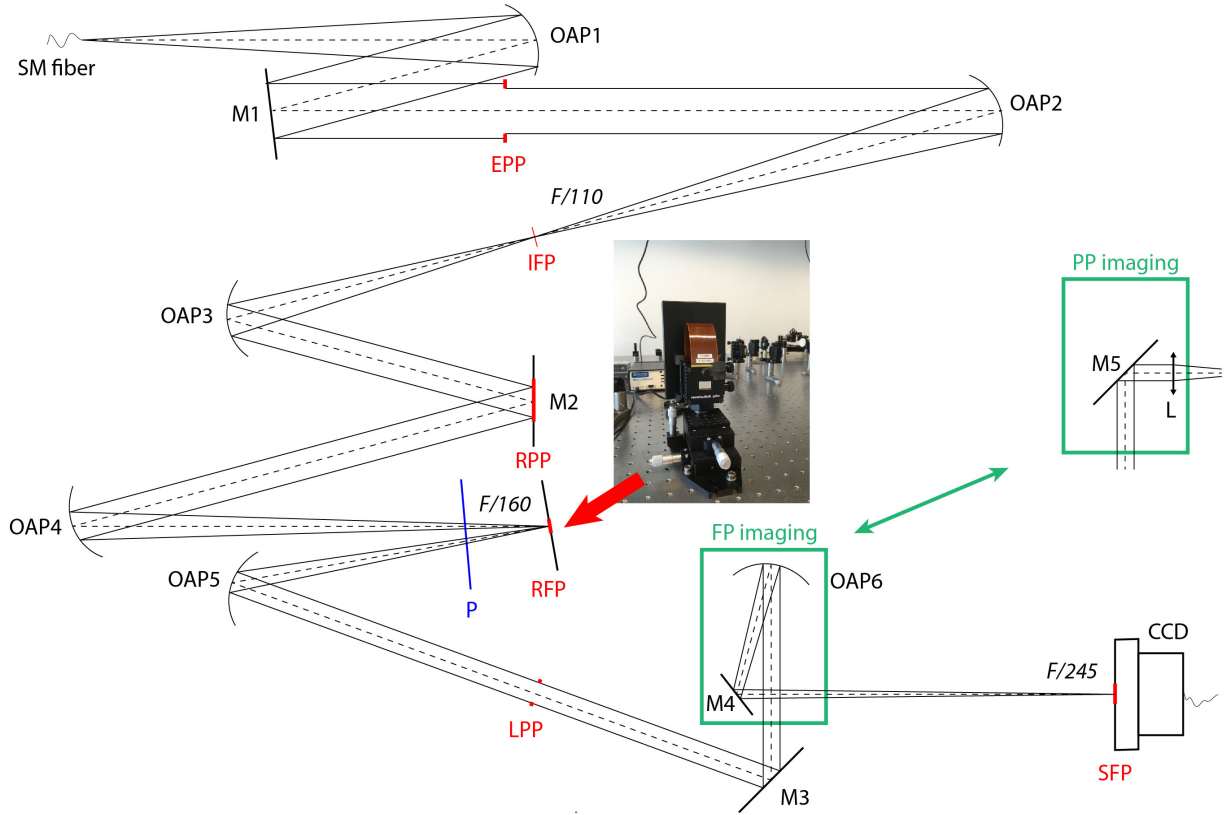

Figure 1. Optical layout of the Swiss Wideband Active Testbed for Coronagraphy and High-contrast imaging (SWATCHi) testbed, dedicated to test LCOS SLM display panels as active focal-plane phase mask coronagraphs in broadband light in the VIS and NIR (with detector upgrade) bands. OAP: off-axis parabolic mirror, M: mirror, P: wire grid polarizer, EPP: entrance pupil-plane, RPP: reflective intermediate pupil-plane, RFP: reflective (coronagraphic) intermediate focal-plane, with SLM (off-axis angle ~5 deg), LPP: Lyot pupil-plane, SFP: scientific focal-plane. The green insets denote switchable focal- (FP) or pupil-plane (PP) imaging stages (interchangeable magnetic plates).

## 2.2 Early broadband results with the Holoeye PLUTO SLM

Over the course of the years 2019-2020, the SWATCHi bench was employed to benchmark coronagraphic nulling performance of various commercially-available LCOS SLM panels from various manufacturers (Holoeye, Meadowlark) in the visible, owning to the absence of a NIR camera over this period. To this purpose, the SWATCHi was operated in three distinct bandpass configurations: (i) monochromatic 633 nm (fiber coupled laser diode, Thorlabs S1FC635), (ii) 6-pc (40 nm) bandwidth centered 640 nm (Semrock FF01-640/40-25 filter), and (iii) 12-pc (75 nm) bandwidth also centered on 640 nm (Semrock FF01-641/75-25 filter). The latter two "broadband" configurations are achieved in conjunction with the use of the supercontinuum source (see description above).

Figure 2 exposes some coronagraphic imaging results obtained with the Holoeye PLUTO-014 LCOS SLM in the three light bandwidth configurations above, and with two entrance aperture geometries: (i) an unobscured free pupil (ideal case), and (ii) a centrally-obscured pupil simulating a ~40 pc obstruction similar to the Palomar Hale or Subaru telescopes (note: this obstructed aperture is about 25 pc smaller than the unobscured one). Two types of coronagraphic FPM phase patterns are programmed onto the SLM: a vortex of topographic charge n = 2, and another vortex with n = 4. As shown with the plots of Figure 3, in the ideal case of the unobstructed aperture, and with both vortices phase masks, the peak-to-peak null depth degrades from ~1.3·10-2 to ~2.2·10-2 when increasing the bandwidth from 0 to 12 pc. Part of this can be explained by the combination of chromatic leakage[10,11] and calibration errors (the SLM was calibrated for

633 nm, not the central wavelength of 640 nm when using the bandpass filters), but combining this[11] with the inherent leakage from the SLM of about $1.3 \cdot 10^{-2}$ – arising mainly from internal parasitic reflections – cannot solely explain this loss in null depth. Still, as exemplified in Fig.3, those effects largely become negligible when dealing with a realistic centrally-obscured pupil, with the null depth remaining essentially flat at $\sim 2 \cdot 10^{-2}$ independently of the bandwidth, with a slight reduction in contrast only noticeable for the $n = 2$ vortex phase map. This is to be expected, and it demonstrates that most sources of leakage from the SLM or from chromatic effects quickly becomes $2^{nd}$-order effects as compared to the impact of non-ideal centrally-obscured apertures and residual AO wavefront errors, typically encountered on most ground-based observatories. These results, as well as similar measurements on a LCOS SLM panel from Meadowlark, are currently planned to be released in the next few months (Kühn et al. 2021, in prep.).

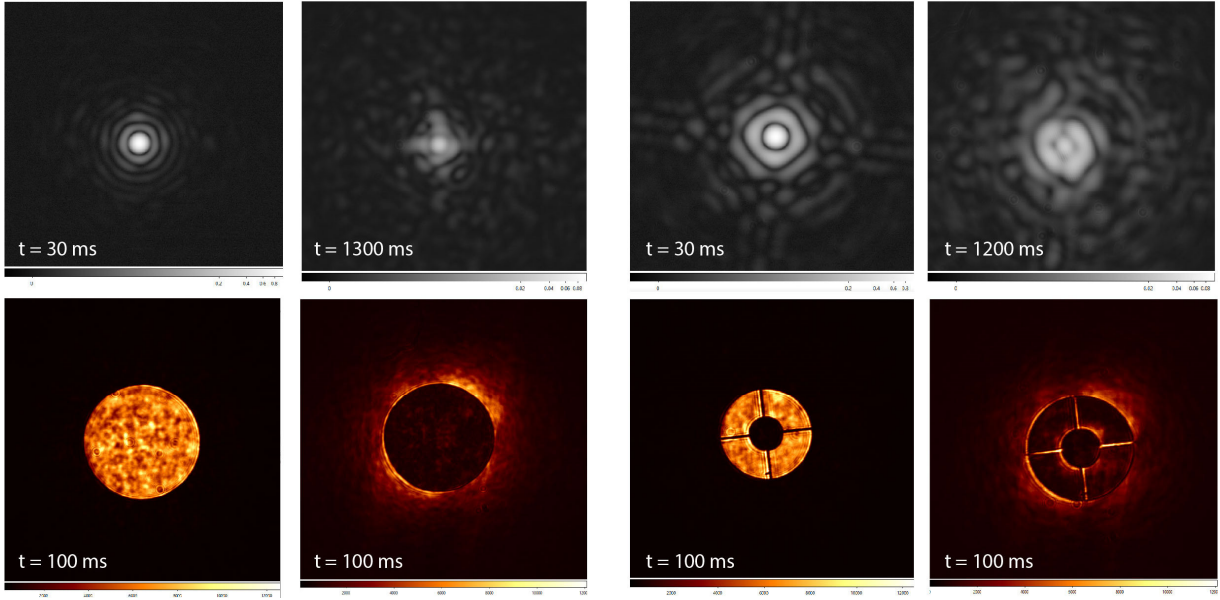

Figure 2. Examples of coronagraphic performance measurements on the SWATCHi bench of Fig.1, with the PLUTO-014 SLM in the RFP plane (see Fig.1) and the 12pc bandwidth filter at 641 nm. (Top row) Non-coronagraphic and coronagraphic PSFs, with a typical broadband peak-to-peak null of $\sim 2.5 \cdot 10^{-2}$; (Bottom row) Respective images of the Lyot pupil plane (LPP on Fig.1), with the Lyot mask taken out to reveal the diffracted star light; (Left panel) Unobscured ideal entrance pupil (EPP on Fig.1) at F/140; (Right panel) Same but with a centrally-obscured entrance pupil (oversized spider struts) at F/180.

**2.3 Upcoming SWATCHi testbed upgrades for H-band operations**

A next logical step, especially in the context of the upcoming PLACID instrument (see Section §3), is to update the SWATCHi bench for NIR operations, in order to qualify NIR SLM panels currently available off the shelves in the telecom band (1.55 μm), very close to the astronomical H-band. Given that all silver-coated optics on SWATCHi are readily compatible with the NIR regime, combined with the fact that the supercontinuum source emits all the way to the astronomical K-band (2.4 μm), this only leaves to upgrade the camera. This is currently foreseen (see §3) to happen around the summer of 2021, with the procurement of an InGaAs camera (C-RED3 from First Light Imaging) that will enable operations up to H-band. At this point, the SWATCHi tesbed will be able to cover the entire wavelength range from 450 nm to 1.65 μm, owing to the procurement of the corresponding bandpass filters.

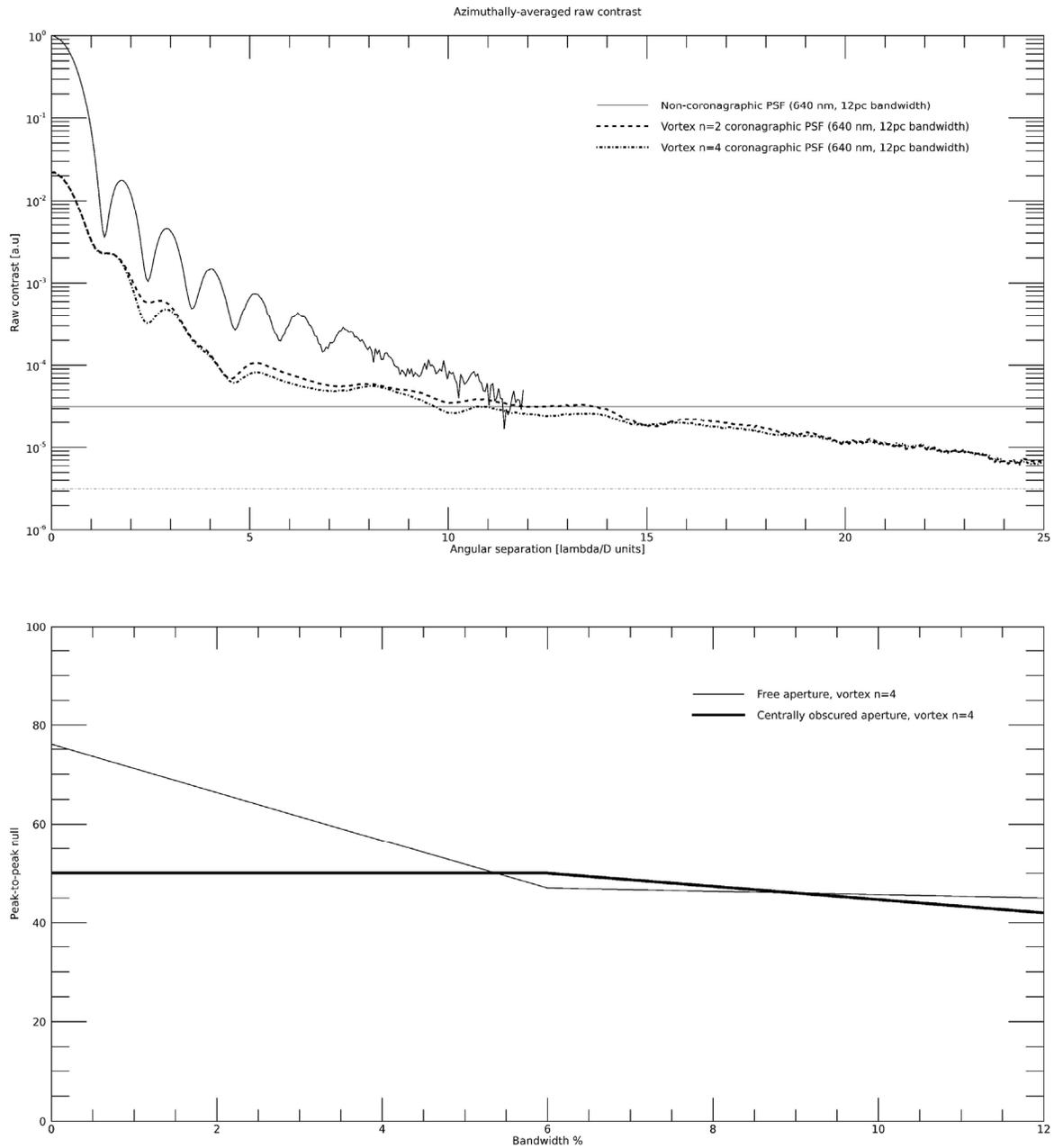

Figure 3. Example of broadband contrast performance for the Holoeye PLUTO-014 SLM with a vortex coronagraph phase pattern. (Top) Raw azimuthally-averaged contrast curves in function of angular separation at 641 nm, 12pc bandwidth, with the horizontal lines denoting the photon noise limits for the non-coronagraphic, respectively the coronagraphic, exposures; (Bottom) Peak-to-peak coronagraphic null in function of bandwidth for the n=4 vortex pattern with an unobscured, respectively a centrally-obscured, entrance aperture.

# 3. THE PLACID INSTRUMENT FOR THE 4-M DAG TELESCOPE

## 3.1 PLACID in a nutshell

The upcoming 4- DAG telescope will be the new national observatory of Turkey, with a Ritchey-Chrétien architecture fielding a 4-m primary mirror and two Nasmyth foci (seeing-limited VIS, and AO-assisted NIR up to 3 μm).[7,8] The DAG observatory will be established in Eastern Anatolia, near Erzurum (site altitude: 3100m, 260/365 clear nights, median seeing ~0.9'') under the leadership of the Atatürk University Astrophysics Research and Application Center (ATASAM, Istanbul), with first light planned by mid-2022. At the time of writing, the dome superstructure and support operation buildings are already completed on-site, with the main telescope structure and primary mirror still sitting in EIE facilities in Italy. Only the AO-assisted Nasmyth platform instrumentation suit is being developed at this stage, and will include the TROIA extreme AO system based on an AlpAO 24x24 deformable mirror and pyramid WFS[9] and a HAWAII-2RG NIR scientific FPA called DIRAC. The DAG telescope will come online at a time where sub-8m observatories are increasingly in need of focusing on niche science cases, but at the same time it will field a state-of-the-art extreme AO system, and its location in the Northern hemispheres still makes its 4-m aperture with Alt-Az configuration highly relevant in the current environment. In this context, in the summer of 2020 ATASAM released a call-for-tender for the procurement of an "adaptive coronagraph" instrument based on the SLM technology, for which a consortium of the University of Bern and the HEIG-VD was officially awarded the contract in November 2020.

The proposed "adaptive coronagraph" instrument is PLACID (Programmable Liquid-crystal Active Coronagraphic Imager for the DAG telescope), and its main characteristics and specifications are summarized on Table 1. PLACID will field a custom NIR SLM from Meadowlark capable of operating from 1.45 μm up to the astronomical Ks-band (2.15 μm), and will be available as an "alternative optical path" fore-optics instrument located in-between the TROIA exAO and the DIRAC FPA. Figure 4 exposes an early-design proposal for PLACID, providing the general design philosophy that will drive the conception of the instrument. One key design driving requirement is to ensure a spatial sampling of at least 10 SLM pixels per λ/D resolution element in the intermediate coronagraphic focal-plane: at H-band this will set the main focal ratio of the instrument at about F/60, which will require several folding optics along the optical path length in order to slow down the F/14 beam from the telescope. In turn, and knowing the SLM panel pixel resolution (1920 x 1152) the F/60 focal ratio will also set the field-of-view (FOV) of PLACID to be strictly larger than 100 x 100 λ/D at H-band, i.e. > 7 x 7 arcsec on sky (see Table 1). Beyond the SLM, the number of active components will be kept minimal, with only a motorized Lyot pupil wheel that will mount various masks optimized for a range of observing conditions or coronagraphic phase maps, and probably a few actuated optics to perform routine alignment. The first-light of PLACID is currently foreseen for the end of 2022.

## 3.2 Brief overview of the road to first-light

The whole PLACID instrument project path to first light is relatively short, with a 24-months period leading to on-site commissioning. This is made possible by the day-one availability of the SWATCHi tesbed to commission the custom NIR SLM from Meadowlark in year 1 in realistic conditions (DAG pupil model) at H-band, with the procurement of a C-RED3 InGaAs "engineering camera", as well as the otherwise passive nature of this intermediate-stage instrument (hence the nickname) that will considerably facilitate optical and mechanical design efforts. This is reinforced by the fact that the HEIG-VD is part of the PLACID consortium, while they are also already in charge of the conception of the TROIA AO system. All in all, the Final Design Review (FDR) of PLACID is foreseen to take place only 12 months into the project timeline, and manufacturing, factory assembly and alignment will take place in the following 9 months, leaving 1 month for factory acceptance and shipping, and 2 months for on-site commissioning.

The electronics and software interface will be relatively straightforward, given that the SLM panels are directly recognized as "2nd display" by a PC graphic card through HDMI connections, and that only a handful of actuated optomechanical components will be required. In practice, this means that the PLACID control box will only integrate a PC with HDMI connection to the SLM and Ethernet link to the control room (remote desktop), as well as a USB hub to operate the various actuated components (off-the-shelves). Further, the AO Nasmyth table will beneficiate from its own temperature- and humidity-regulated enclosure, preventing the need to develop temperature control elements for the SLM and the PLACID bench.

Table 1. Overview of PLACID main specifications (as of December 2020).

| Specifications | Design value |
|---|---|
| Operating wavelength | H-band $\lambda_0$ = 1.6 µm (baseline) |
| | Ks-band $\lambda_0$ = 2.15 µm (best effort) |
| Active coronagraph SLM | Custom NIR 1920x1152 SLM (Meadowlark) |
| | Refresh rate : 60 Hz; Pixel pitch : 9.2 µm |
| Focal ratio | ~F/60 (to warrant 10 SLM pixel per $\lambda/D$) |
| On-sky FOV | > 7 x 7 arcsec |
| Coronagraphic null depth | < $10^{-1}$ with internal light source |
| Inner-working angle | < 2.5 $\lambda/D$ with internal light source |
| Optical throughput | > 10% |
| Interfacing | HDMI (SLM) + USB (actuated opto-mechanics) |
| Future upgrade options | • NCPA WFS (Zernike WFS) |
| | • Binary/multiple stars ADI imaging |
| | • Time-domain CDI observations |

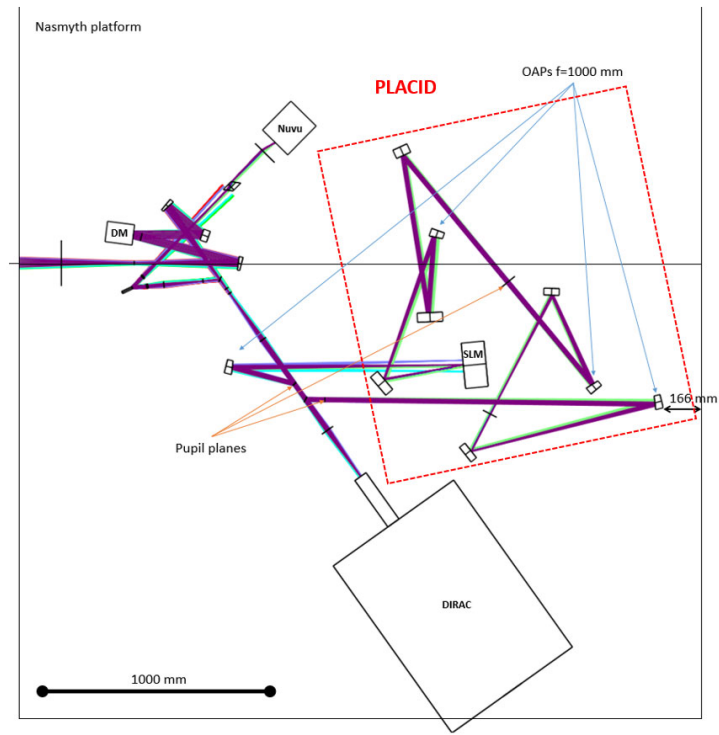

Figure 4. Early PLACID optical design concept. PLACID will be a standalone optional (actuated flip mirrors) ~F/60 optical path on the AO-assisted Nasmyth platform of the DAG telescope, which will be transparent to the DIRAC NIR FPA as compared to the non-coronagraphy path at F/14. An actuated filter wheel will be inserted in the denoted post-coronagraphic pupil-plane, to be able to switch from various Lyot masks depending on which phase pattern is programmed onto the SLM.

### 3.3 Foreseen first-light capabilities and future upgrades path

From day one, PLACID will provide the DAG telescope with adaptive coronagraphic capabilities at H- and Ks-band, enabling AO-assisted high-contrast imaging of exoplanets and circumstellar material. By "adaptive coronagraphy it is meant that PLACID will be able to operate with various FPM coronagraphic schemes with a simple mouse click, among which vortex coronagraphy (any topographic charge), four-quadrant or eight-octant coronagraphy, Zernike/Roddier&Roddier coronagraphy, or any user-defined focal-plane phase map. Beyond the ability to experiment with a variety of patterns for a given science target, it is foreseen that this versatility will considerably help to adapt to changes in observing conditions, whether by being able to digitally shift the phase pattern in x- and y-directions, but particularly with the ability to tune the "aggressiveness" (e.g. the topographic charge) of a given coronagraph in real-time, for example trading inner-working angle (IWA) for improved robustness vs. tip-tilt jitter in case of deteriorating wind conditions. It is expected that in turn the DAG high-contrast imaging instrument mode will benefit from improved on-sky observing efficiency over the time, being less constrained by seeing or weather conditions, as opposed to equivalent high-contrast imager on ground-based telescopes. This feature, combined with the ease-of-use for the night operator ("all digital" operations) and the competitiveness of the PLACID DAG platform for international collaborations (on-sky coronagraphic laboratory for prototyping new proof-of-concepts FPMs with a simple file upload), is anticipated to provide the DAG observatory with some unique niche capabilities in spite of its modest aperture.

Moreover, looking forward and over the next few years after first-light (2023 and beyond), there will be a wealth of new capabilities and/or upgrade routes to consider for PLACID, which are deemed to provide the nascent Turkish astronomy community with a variety of R&D and PhD projects for instrumentalists. Some of those upgrades will mainly consist in software-only developments, among which NPCA calibrations with the phase-shifting Zernike scheme,[12] or nulling of multiple stars to enable high-contrast imaging of challenging compact of binary (or higher multiple) systems, with compatibility with angular differential imaging through digital real-time rotation of the phase pattern.[1] Additional upgrade paths include faster SLMs (600-700 Hz), or even the prototyping of polarization-insensitive SLM displays.[13] A particularly compelling long-term research pathway would consist in combining the use of fast SLMs with high-speed low-noise NIR scientific FPAs, such as MKIDS,[14] to be able to freeze the atmospheric speckles and employ time-domain modulation to implement coherent differential imaging.[3,15]

## 4. CONCLUSIONS

Overall, SLM-based adaptive coronagraphy admittedly only provides modest contrast, as well as particularly throughput, performance at the moment, mainly due to technological limitations. Some of those limitations might be partially overcome in the next few years, if sufficient effort is invested in developing and testing polarization-independent SLM panels,[13] or by validating angular-dispersion compensation optical schemes while using SLM panels in "diffraction grating" mode to minimize sensitivity to parasitic reflections. Nevertheless, although some weaknesses might be partially compensated by smart use of the versatility provided by the approach, it is to be expected that global contrast and sensitivity performance will keep trailing those of "classical coronagraphy" for quite some time to come, at least in the laboratory. Indeed, with the notable exception of space-based observation platforms, most ground-based high-contrast instruments are generally limited in performance by the "bottleneck" of post-AO residual wavefront errors, and this is deemed to stay true for a while.

In this context, the advent of the PLACID instrument on a AO-equipped ($2^{nd}$-generation, pyramid wavefront sensor) 4-m class telescope such as the DAG observatory, in a few years from now, is foreseen to provide a visibility boost and multiple instrumentation research paths to nurture a still-nascent Turkish astronomy community. This will be combined with a versatile high-contrast imaging platform, possibly with improved observational efficiency vs. weather conditions as compared to the competition, available to all Turkish observers from day one. It is further expected that up to 50% of the DAG observing time will be made open to the international community through open calls, which will reinforce the attractiveness of the platform for on-sky research and development in coronagraphic instrumentation.

Most importantly, beyond the obvious niche position in high-contrast imaging instrumentation, the location of the DAG telescope in the Northern hemisphere, the expected extreme AO performance delivered by the TROIA facility, and the foreseen wealth of available observing nights in the first few years following the DAG first-light, mainly a consequence of the limited initially-available instrumentation suit, may very well ensure a valuable science pathway for PLACID. To name a few, one could make strong use of the available observing time to setup systematic follow-up programs of large transit missions, such as TESS and PLATO, or rely on the niche PLACID features to be able to survey a large population of binary or multiple-star systems in high-contrast imaging mode, in the wake of recent efforts undertaken with the Stellar Double Coronagraph (SDC) instrument on the Palomar Hale Telescope.[16]